\documentclass[preprint2]{aastex6}
\usepackage{amsmath}
\usepackage{booktabs}
\usepackage{longtable}
\usepackage{xspace}
\usepackage{hyperref}
\usepackage[T1]{fontenc}
\usepackage{lipsum}

\newcommand{\Kepler}{\textit{Kepler}\xspace}

\newcommand{\isochrones}{\texttt{isochrones}\xspace}
\newcommand{\isoclassify}{\texttt{isoclassify}\xspace}

\newcommand{\Gaia}{\textit{Gaia}\xspace}

\newcommand{\rchisq}{\ensuremath{\chi_r^2}\xspace}

\newcommand{\Mstar}{\ensuremath{M_{\star}}\xspace}
\newcommand{\Rstar}{\ensuremath{R_{\star}}\xspace} 
 
\newcommand{\teff}{\ensuremath{T_{\mathrm{eff}}}\xspace}  
\newcommand{\logg}{\ensuremath{\log g}\xspace} 
\newcommand{\fe}{[Fe/H]\xspace}
\newcommand{\vsini}{\ensuremath{v \sin i}\xspace} 

\newcommand{\logage}{\ensuremath{\log_{10}{(\mathrm{age})}}\xspace}
\newcommand{\plxiso}{\ensuremath{\pi_{\star,\mathrm{iso}}}\xspace}
\newcommand{\plxtrig}{\ensuremath{\pi_{\star,\mathrm{trig}}}\xspace}
\newcommand{\teffiso}{\ensuremath{T_{\mathrm{eff,iso}}}\xspace}
\newcommand{\loggiso}{\ensuremath{\log g_\mathrm{iso}}\xspace} 
\newcommand{\feiso}{\ensuremath{\mathrm{[Fe/H]}_\mathrm{iso}}\xspace}

\newcommand{\Rp}{\ensuremath{R_P}\xspace}
\newcommand{\teq}{\ensuremath{T_{\mathrm{eq}}}\xspace}
\newcommand{\Sinc}{\ensuremath{S_{\mathrm{inc}}}\xspace}

\renewcommand{\Re}{\ensuremath{R_{\oplus}}\xspace}

\newcommand{\Rsun}{\ensuremath{R_{\odot}}\xspace }
\newcommand{\Msun}{\ensuremath{M_{\odot}}\xspace}

\def\deg{\ensuremath{^{\circ}}}

\usepackage{xstring} 

\newcommand{\val}[1]{%
  \IfEqCase{#1}{%
{nstars-cks}{1305}
{ncand-cks}{2025}
{equad-smass}{2\%}
{equad-srad-dw}{2\%}
{equad-srad-gi}{10\%}
{equad-sage}{10\%}
{iso-smass-err-05}{2.6\%}
{iso-smass-err-50}{3.3\%}
{iso-smass-err-95}{7.5\%}
{iso-srad-err-05}{3.1\%}
{iso-srad-err-50}{10.7\%}
{iso-srad-err-95}{16.6\%}
{iso-slogage-err-05}{0.05~dex}
{iso-slogage-err-50}{0.10~dex}
{iso-slogage-err-95}{0.30~dex}
{iso-floor-smass-err-05}{3.3\%}
{iso-floor-smass-err-50}{3.9\%}
{iso-floor-smass-err-95}{7.8\%}
{iso-floor-srad-err-05}{3.7\%}
{iso-floor-srad-err-50}{11.0\%}
{iso-floor-srad-err-95}{17.3\%}
{iso-floor-slogage-err-05}{0.07~dex}
{iso-floor-slogage-err-50}{0.11~dex}
{iso-floor-slogage-err-95}{0.30~dex}
{cks-huber-nstars}{72}
{cks-huber-smass-ratio-mean}{98.3\%}
{cks-huber-smass-ratio-std}{6.3\%}
{cks-huber-srad-ratio-mean}{95.2\%}
{cks-huber-srad-ratio-std}{11.2\%}
{cks-huber-dwarf-srad-ratio-mean}{95.8\%}
{cks-huber-dwarf-srad-ratio-std}{9.7\%}
{cks-tgas-sparallax-diff-mean}{0.14}
{cks-tgas-sparallax-diff-std}{0.51}
{cks-silva-nstars}{32}
{cks-silva-slogage-diff-mean}{0.02~dex}
{cks-silva-slogage-diff-std}{0.10~dex}
{huberphot-nstars-tot}{1277}
{huberphot-nstars-phot}{969}
{huberphot-nstars-ast}{88}
{huberphot-nstars-spec}{220}
{huberphot-srad-frac-err-median}{39\%}
{huberphot-srad-ast-frac-err-median}{2.9\%}
{huberphot-srad-phot-frac-err-median}{42\%}
{huberphot-srad-spec-frac-err-median}{17\%}
{huberphot-smass-frac-err-median}{14\%}
{huberphot-smass-ast-frac-err-median}{6.9\%}
{huberphot-smass-phot-frac-err-median}{16\%}
{huberphot-smass-spec-frac-err-median}{7.1\%}
{cks-huberphot-smass-noutliers}{9}
{cks-huberphot-smass-ratio-mean}{1.8\%}
{cks-huberphot-smass-ratio-std}{11.2\%}
{cks-huberphot-srad-noutliers}{7}
{cks-huberphot-srad-ratio-mean}{2.8\%}
{cks-huberphot-srad-ratio-std}{28.2\%}
{cks-rp-frac-err-median}{12\%}
{cks-sinc-frac-err-median}{21\%}
{cks-sma-frac-err-median}{1.7\%}
{nstars-nea}{2025}
{nea-rp-frac-err-median}{38\%}
{nea-sinc-frac-err-median}{113\%}
{nea-srad-frac-err-median}{38\%}
{nea-smass-frac-err-median}{13.4\%}
{nea-rp-frac-err-median-ast}{2.8\%}
{nea-rp-frac-err-median-spec}{17\%}
{nea-rp-frac-err-median-phot}{42\%}
{nea-sinc-frac-err-median-ast}{12\%}
{nea-sinc-frac-err-median-spec}{48\%}
{nea-sinc-frac-err-median-phot}{124\%}
{cks-stars-with-furlan-rcorr-gt-5percent}{141}
{cks-huber-smass-offset}{1.7\% smaller} 
{cks-huber-dwarf-srad-offset}{4.8\% smaller} 
{cks-silva-slogage-diff-offset}{0.02~dex larger} 
{cks-tgas-sparallax-diff-offset}{0.16~mas larger} 
{cks-huber-srad-rchisq}{1.0}
{cks-huber-smass-rchisq}{0.6}
{cks-silva-slogage-rchisq}{0.6}
{cks-smass-ferr-round}{4\%} 
{cks-srad-ferr-round}{11\%}
{cks-sage-ferr-round}{30\%} 
{cks-prad-ferr-round}{12\%}
{cks-sinc-ferr-round}{21\%}
}[\PackageError{tree}{Undefined option to tree: #1}{}]%
}%

\submitted{Submitted to AAS Journals}
\begin{document}

\title{The California-Kepler Survey. \\
      II. Precise Physical Properties of 2025 Kepler Planets and Their Host Stars\altaffilmark{1}}
\author{
John Asher Johnson\altaffilmark{2},
Erik A.\ Petigura\altaffilmark{3,9,14},
Benjamin J.\ Fulton\altaffilmark{4,11},
Geoffrey W.\ Marcy\altaffilmark{5},
Andrew W.\ Howard\altaffilmark{3,4}, 
Howard Isaacson\altaffilmark{5},
Leslie Hebb\altaffilmark{6}, 
Phillip A.\ Cargile\altaffilmark{2},
Timothy D.\ Morton\altaffilmark{8},
Lauren M.\ Weiss\altaffilmark{7,12}, 
Joshua N.\ Winn\altaffilmark{8}, 
Leslie A.\ Rogers\altaffilmark{9}, 
Evan Sinukoff\altaffilmark{3,2,13}, 
Lea A.\ Hirsch\altaffilmark{4} 
}

\altaffiltext{1}{Based on observations obtained at the W.\,M.\,Keck Observatory, which is operated jointly by the University of California and the California Institute of Technology. Keck time has been granted by the University of California, and California Institute of Technology, the University of Hawaii, and NASA.} 
\altaffiltext{2}{Harvard-Smithsonian Center for Astrophysics, 60 Garden St, Cambridge, MA 02138, USA}
\altaffiltext{3}{California Institute of Technology, Pasadena, CA, 91125, USA}
\altaffiltext{4}{Institute for Astronomy, University of Hawai`i at M\={a}noa, Honolulu, HI 96822, USA}
\altaffiltext{5}{Department of Astronomy, University of California, Berkeley, CA 94720, USA}
\altaffiltext{6}{Hobart and William Smith Colleges, Geneva, NY 14456, USA}
\altaffiltext{7}{Institut de Recherche sur les Exoplan\`{e}tes, Universit\'{e} de Montr\'{e}al, Montr\'{e}al, QC, Canada}
\altaffiltext{8}{Department of Astrophysical Sciences, Peyton Hall, 4 Ivy Lane, Princeton, NJ 08540 USA}
\altaffiltext{9}{Department of Astronomy \& Astrophysics, University of Chicago, 5640 South Ellis Avenue, Chicago, IL 60637, USA}
\altaffiltext{10}{Hubble Fellow}
\altaffiltext{11}{National Science Foundation Graduate Research Fellow}
\altaffiltext{12}{Trottier Fellow}
\altaffiltext{13}{Natural Sciences and Engineering Research Council of Canada Graduate Student Fellow}
\altaffiltext{14}{Corresponding author: petigura@caltech.edu}

\begin{abstract}
We present stellar and planetary properties for $\val{nstars-cks}$ Kepler Objects of Interest (KOIs) hosting $\val{ncand-cks}$ planet candidates observed as part of the California-Kepler Survey. We combine spectroscopic constraints, presented in Paper I, with stellar interior modeling to estimate stellar masses, radii, and ages. Stellar radii are typically constrained to \val{cks-srad-ferr-round}, compared to 40\% when only photometric constraints are used. Stellar masses are constrained to \val{cks-smass-ferr-round}, and ages are constrained to \val{cks-sage-ferr-round}. We verify the integrity of the stellar parameters through comparisons with asteroseismic studies and \Gaia parallaxes. We also recompute planetary radii for $\val{ncand-cks}$ planet candidates. Because knowledge of planetary radii is often limited by uncertainties in stellar size, we improve the uncertainties in planet radii from typically \val{nea-rp-frac-err-median-phot} to \val{cks-prad-ferr-round}. We also leverage improved knowledge of stellar effective temperature to recompute incident stellar fluxes for the planets, now precise to \val{cks-sinc-ferr-round}, compared to a factor of two when derived from photometry.
\end{abstract}

\section{Introduction}
The prime \Kepler mission (2009--2013; \citealt{Borucki10a}) revealed over 4000 planet candidates \citep{Mullally15}. The vast majority of these planet candidates, formally known as Kepler Objects of Interest (KOIs), are {\em bona fide} planets \citep{Morton11,Lissauer12}. This large sample of planets with high purity enabled studies of planet occurrence \citep{Howard12,Fressin13,Petigura13b} and planetary architectures \citep{Lissauer11,Fabrycky14}, and when coupled with spectroscopy, enabled determination of planet masses, densities, and interiors \citep{Marcy2014,Weiss2014,Rogers2015,Wolfgang2015}. However, the inferred properties of extrasolar planets are often limited by uncertainties in stellar properties. The Kepler Input Catalog \citep[KIC;][]{Brown11} was the first homogeneous catalog of properties of \Kepler field stars. However, stellar radii (\Rstar) in the KIC, based solely on photometric constraints, have fractional uncertainties of $\sigma(\Rstar) /\Rstar \approx 40\%$, which limits the precision with which one can measure planetary radii and densities.

The California-Kepler Survey (CKS) is a large spectroscopic survey conducted with Keck/HIRES of KOIs. This survey was conducted with the aim of improving knowledge of host star properties, which translate into higher precision measurements of planetary properties including planet radius (\Rp) and incident stellar flux (\Sinc). The CKS project and goals are described in detail in Paper I of this series (Petigura et al. 2017). In brief, between 2012 and 2015 we obtained high-resolution ($R \approx 50,000$) spectra of $\val{nstars-cks}$ stars identified as KOIs with Keck/HIRES \citep{Vogt94}. We used an exposure meter to achieve a uniform signal-to-noise ratio $\approx45$ per HIRES pixel on blaze near 5500~\AA. Using these spectra, we derived effective temperature (\teff), surface gravity (\logg), metallicity (\fe), and projected stellar rotation velocity (\vsini).

In this work (Paper II of the CKS series), we convert the observed spectroscopic properties of Paper I into physical stellar and planetary properties. In Section~\ref{sec:stellar}, we convert \teff, \logg, and \fe into stellar masses, radii, and ages. We assess the integrity of these measurements through comparisons with asteroseismology and trigonometric parallaxes from \Gaia. We find that the typical fractional uncertainties in \Mstar and \Rstar are \val{cks-smass-ferr-round} and \val{cks-srad-ferr-round}, respectively. Stellar ages are constrained to \val{cks-sage-ferr-round}. In Section~\ref{sec:planet}, we recompute planetary parameters including \Rp and \Sinc. We offer some concluding thoughts in Section~\ref{sec:conclusion} and introduce subsequent papers in the CKS series that leverage these improved stellar and planetary properties.

\section{Stellar Properties}
\label{sec:stellar}
\subsection{Isochrone Modeling}
\label{ssec:isochrone}
Several groups have used theoretical models of stellar structure and evolution to compile grids of stellar properties (\Rstar, \teff, etc.) as function of \Mstar, \fe, and age. A set of models at constant metallicity and age is commonly called an ``isochrone.'' We used the Dartmouth Stellar Evolution Program (DSEP) models \citep{Dotter08} to convert the spectroscopic properties of \teff, \logg, and \fe into \Mstar, \Rstar, and age. To facilitate this conversion, we used the publicly-available Python package \isochrones \citep{Morton15},%
\footnote{\url{https://github.com/timothydmorton/isochrones} (version 1.0)}
which interpolates between the discrete grid of DSEP models to derive properties at off-grid values. 

One feature of \isochrones is the capability to use Markov Chain Monte Carlo (MCMC) sampling%
\footnote{Specifically, the affine-invariant ensemble sampler of \cite{Goodman10}, as implemented in Python by \cite{Foreman-Mackey13}}
to compute the range of physical parameters (\Mstar, \Rstar, age, and other parameters), consistent with a set of user-defined observational constraints. In order to facilitate convergence, we seeded the sampler with initial guesses of \Mstar and age, which we computed using the publicly-available Python package \isoclassify \citep{Huber17},%
\footnote{https://github.com/danxhuber/isoclassify}
which uses the MESA Isochrones and Stellar Tracks (MIST) database \citep{Choi16,Paxton11,Paxton13,Paxton15}. Because \isoclassify is a grid-based code, initializing \isochrones takes only a few CPU-seconds per star. We then performed the more computationally-expensive MCMC exploration of the likelihood surface using \isochrones, which requires several CPU-minutes per star.

For each star, \isochrones returned the set of stellar masses, radii and ages consistent with the spectroscopic \teff, \logg, and \fe from Paper~I. While the stellar parameters of interest could be derived solely from the spectroscopic parameters, we included a single photometric band in order to estimate distances and to facilitate a comparison with {\em Gaia} parallaxes (Section~\ref{sec:gaia}). We used {\em K}-band from 2MASS \citep{2MASS06}  because it was the reddest band available and thus least sensitive to interstellar extinction.

We elected against including broadband photometry from multiple bands because doing so could have the deleterious effect of biasing our results away from the spectroscopic values. For \Kepler target stars, typical uncertainties for 2MASS $K$ and $J$ apparent magnitudes are 0.02~mag. However, for a G2 star, an error in $J - K$ color of 0.02~mag corresponds difference in \teff of $\approx100$~K, which is larger than the 60~K precision of the CKS effective temperatures (\citealt{Casagrande10,Pecaut13}). The potential for such biases is compounded by uncertain line-of-sight extinction and photometric zero-point errors.  We thus used only a single photometric band to avoid such biases. Interstellar extinction or zero-point errors in the input {\em K}-band magnitudes could influence the implied source distance, but not the derived \Mstar, \Rstar, and age, which are constrained solely from spectroscopy.

We list our derived \Mstar, \Rstar, and age measurements and uncertainties in Table~\ref{tab:stellar-parameters}, which are computed from the 16, 50, and 84 percentiles of the posterior samples. During the modeling, \isochrones also samples \teff, \logg, and \fe. Typically, these parameters reflect the input \teff, \logg, and \fe from spectroscopy with our adopted uncertainties. In some cases, where, by fluctuations or other errors, the spectroscopic constraints extend into regions of the HR diagram that are not populated by the DSEP models. In these cases, \isochrones only samples \teff, \logg, \fe, etc that are allowed by the physics incorporated in the DSEP models. The behavior occurs most often in cool dwarf stars  ($\teff \lesssim 5300$~K) where the main sequence has a narrow spread in \logg. Following the notation of \cite{Valenti05}, we also list these isochrone-constrained properties, \teffiso, \loggiso, \feiso in Table~\ref{tab:stellar-parameters}.

The DSEP models also tabulate absolute stellar magnitudes in various band-passes. By comparing stellar apparent magnitude to the theoretical absolute magnitude, one can compute an ``isochrone parallax,'' modulo line-of-sight extinction to the target star. In Table~\ref{tab:stellar-parameters}, we list this implied parallax, which we denote \plxiso, to distinguish from trigonometric parallax, \plxtrig. We perform a comparison of \plxiso and \plxtrig in Section~\ref{sec:gaia}.

\begin{deluxetable*}{llRRRRRRRRR}
\tablecaption{Stellar Properties\label{tab:stellar-parameters}}
\tabletypesize{\scriptsize}
\tablecolumns{11}
\tablewidth{0pt}
\tablehead{
	\colhead{KOI} & 
	\colhead{Tycho-2} & 
	\colhead{{\em K}} &
	\colhead{\teffiso} &
	\colhead{\loggiso} & 
	\colhead{\feiso} &
	\colhead{\Mstar} & 
    \colhead{\Rstar} & 
    \colhead{\logage} &
    \colhead{\plxiso} &
    \colhead{\plxtrig}
    \\
    \colhead{} & 
    \colhead{} & 
    \colhead{mag} & 
	\colhead{K} &
	\colhead{dex} & 
	\colhead{dex} &
	\colhead{\Msun} & 
    \colhead{\Rsun} & 
    \colhead{} &
    \colhead{mas} &
    \colhead{mas} 
}
\startdata
K00001 & TYC 3549-2811-1 & 9.8  & 5815_{ -65 }^{ +66 }  & 4.39_{ -0.09 }^{ +0.08 }  & 0.01_{ -0.04 }^{ +0.04 }  & 1.01_{ -0.03 }^{ +0.03 }  & 1.06_{ -0.09 }^{ +0.12 }  & 9.75_{ -0.32 }^{ +0.13 }  & 4.68_{ -0.49 }^{ +0.43 }  & 4.32_{ -0.25 }^{ +0.25 } \\
K00002 & TYC 3547-1402-1 & 9.3  & 6447_{ -64 }^{ +65 }  & 4.15_{ -0.11 }^{ +0.11 }  & 0.19_{ -0.04 }^{ +0.04 }  & 1.37_{ -0.07 }^{ +0.09 }  & 1.63_{ -0.22 }^{ +0.27 }  & 9.36_{ -0.12 }^{ +0.07 }  & 3.63_{ -0.51 }^{ +0.56 }  & 2.99_{ -0.42 }^{ +0.42 } \\
K00003 & \nodata & 7.0  & 4867_{ -65 }^{ +66 }  & 4.54_{ -0.03 }^{ +0.04 }  & 0.31_{ -0.04 }^{ +0.04 }  & 0.83_{ -0.03 }^{ +0.03 }  & 0.81_{ -0.03 }^{ +0.03 }  & 9.97_{ -0.39 }^{ +0.16 }  & 25.30_{ -1.10 }^{ +1.10 }  & \nodata\\
K00006 & TYC 3135-372-1 & 11.0  & 6344_{ -67 }^{ +65 }  & 4.32_{ -0.08 }^{ +0.06 }  & 0.04_{ -0.04 }^{ +0.04 }  & 1.23_{ -0.04 }^{ +0.04 }  & 1.26_{ -0.09 }^{ +0.14 }  & 9.29_{ -0.39 }^{ +0.17 }  & 2.19_{ -0.22 }^{ +0.17 }  & 2.43_{ -0.33 }^{ +0.33 } \\
K00007 & \nodata & 10.8  & 5833_{ -67 }^{ +60 }  & 4.12_{ -0.10 }^{ +0.11 }  & 0.17_{ -0.04 }^{ +0.04 }  & 1.12_{ -0.06 }^{ +0.10 }  & 1.53_{ -0.20 }^{ +0.24 }  & 9.81_{ -0.13 }^{ +0.09 }  & 2.08_{ -0.28 }^{ +0.32 }  & \nodata\\
K00008 & \nodata & 11.0  & 5883_{ -64 }^{ +67 }  & 4.46_{ -0.07 }^{ +0.04 }  & -0.06_{ -0.04 }^{ +0.04 }  & 1.02_{ -0.04 }^{ +0.04 }  & 0.98_{ -0.05 }^{ +0.07 }  & 9.47_{ -0.47 }^{ +0.25 }  & 2.89_{ -0.20 }^{ +0.17 }  & \nodata\\
K00010 & \nodata & 12.3  & 6179_{ -63 }^{ +68 }  & 4.25_{ -0.11 }^{ +0.09 }  & -0.07_{ -0.04 }^{ +0.05 }  & 1.13_{ -0.05 }^{ +0.07 }  & 1.31_{ -0.14 }^{ +0.22 }  & 9.60_{ -0.11 }^{ +0.08 }  & 1.19_{ -0.17 }^{ +0.14 }  & \nodata\\
K00017 & \nodata & 11.6  & 5667_{ -63 }^{ +58 }  & 4.17_{ -0.10 }^{ +0.10 }  & 0.34_{ -0.04 }^{ +0.04 }  & 1.11_{ -0.06 }^{ +0.10 }  & 1.45_{ -0.18 }^{ +0.22 }  & 9.83_{ -0.13 }^{ +0.10 }  & 1.53_{ -0.21 }^{ +0.22 }  & \nodata\\
K00018 & \nodata & 11.8  & 6333_{ -69 }^{ +67 }  & 4.15_{ -0.11 }^{ +0.10 }  & 0.03_{ -0.04 }^{ +0.04 }  & 1.29_{ -0.07 }^{ +0.09 }  & 1.59_{ -0.21 }^{ +0.28 }  & 9.47_{ -0.07 }^{ +0.06 }  & 1.22_{ -0.18 }^{ +0.18 }  & \nodata\\
K00020 & \nodata & 12.1  & 5927_{ -65 }^{ +65 }  & 4.05_{ -0.10 }^{ +0.10 }  & 0.03_{ -0.04 }^{ +0.04 }  & 1.14_{ -0.06 }^{ +0.09 }  & 1.67_{ -0.22 }^{ +0.26 }  & 9.77_{ -0.11 }^{ +0.08 }  & 1.06_{ -0.14 }^{ +0.16 }  & \nodata\\

\enddata
\tablecomments{Stellar parameters for the $\val{nstars-cks}$ stars in the California-Kepler Survey (CKS) catalog. We provide the Tycho-2 identifier, where available. {\em K} is the apparent {\em K}-band magnitude from the Two Micron All Sky Survey (2MASS, \citealt{2MASS06}). We used the \isochrones Python package to derive the following physical parameters: \teffiso, \loggiso, \feiso, \Mstar, \Rstar, \logage, and \plxiso. \isochrones returns posterior distributions on effective temperature, surface gravity, and metallicity, which we distinguish from the purely spectroscopic measurements as \teffiso, \loggiso, \feiso. We list the trigonometric parallax (\plxtrig) for stars listed in the Tycho-Gaia Astrometric Solution (TGAS). Table \ref{tab:stellar-parameters} is published in its entirety in machine-readable format. A portion is shown here for guidance regarding its form and content.}
\end{deluxetable*}

\subsection{Uncertainties}
\label{ssec:uncertainties}
The \isochrones framework computes the range of physical parameters (\Mstar, \Rstar, age, etc), consistent with the spectroscopic input constraints. The formal uncertainties on the stellar parameters are set by the uncertainties associated with \teff, \logg, and \fe. The median formal uncertainties are \val{iso-smass-err-50}, \val{iso-srad-err-50}, and \val{iso-slogage-err-50} in \Mstar, \Rstar, and age, respectively. The formal uncertainties associated with an individual star depend on its position in the HR diagram. We list the ranges of these uncertainties in Table~\ref{tab:uncert-formal}. For \Mstar and \Rstar, the fractional errors are smallest for cool ($\teff \lesssim 5300$~K) main-sequence stars because these quantities are constrained mainly by \teff and \fe, which are known to high precision from spectroscopy. There is little dispersion in the DSEP models as function of stellar age, due to the long timescales associated with the evolution of low-mass stars. Consequently, the formal uncertainties on stellar age are largest for these low-mass stars. The formal uncertainties on \Mstar and \Rstar are largest for evolved stars, because a larger variety of (\Mstar, \Rstar, age) combinations are consistent with the spectroscopic observables.

The formal errors from \isochrones do not incorporate model-dependent uncertainties associated with the DSEP models. Quantifying the extent to which stellar models can accurately reproduce the physical properties of real stars involves detailed comparisons with stars that have physical parameters measured through some model-independent means such as eclipsing binary systems or interferometry (e.g. \citealt{Boyajian12}). Such comparisons are an active area of research and are beyond the scope of this paper. 

Here, we make an estimate of the size of such errors by comparing the physical parameters derived using two sets of models. In addition to interpolating between the DSEP tracks, \isochrones can also interpolate between models from MESA Isochrones and Stellar Tracks (MIST) database (\citealt{Choi16,Paxton11,Paxton13,Paxton15}). We performed a parallel analysis of our spectroscopic parameters using the MIST models and compared the derived parameters. We found that \Mstar and age computed using these methods were consistent to 2\% and 10\%, respectively. For \Rstar, the degree of agreement between the models depended on whether a star had evolved off the main-sequence. The radii were consistent to 2\% and 10\% for dwarfs and evolved stars, respectively. In this paper, we have adopted \logg = 3.9~dex as a convenient dividing line between between dwarfs and evolved stars.

To account for model-dependent uncertainties, we added the following fractional uncertainties in quadrature to our CKS parameters: We added 2\% to the mass uncertainties; we added 2\% and 10\% to the radius uncertainties for dwarfs and evolved stars respectively; and we added 10\% to the age uncertainties.

Our adopted uncertainties are summarized in Table~\ref{tab:uncert-formal}. Because the model-dependent uncertainties are typically smaller than the formal uncertainties, the inclusion of model-dependent errors affects only a small number of stars. 

\begin{deluxetable*}{lrrrrrrrr}
\tablecaption{Summary of Parameter Uncertainties\label{tab:uncert-formal}}
\tablecolumns{8}
\tablewidth{0pt}
\tablehead{
\colhead{} &	
\multicolumn{3}{c}{Formal}  &
\colhead{} &
\multicolumn{3}{c}{Adopted}  \\
\cline{2-4} 
\cline{6-8} \\[-3.5ex]
\colhead{Parameter} & 
\colhead{5\%} &
\colhead{50\%} & 
\colhead{95\%} &
\colhead{} &
\colhead{5\%} &
\colhead{50\%} & 
\colhead{95\%} 
}
\startdata
$\sigma(\Mstar)/\Mstar$ & \val{iso-smass-err-05} & \val{iso-smass-err-50} & \val{iso-smass-err-95}  & ~ & \val{iso-floor-smass-err-05}   & \val{iso-floor-smass-err-50} & \val{iso-floor-smass-err-95}  \\ 
$\sigma(\Rstar)/\Rstar$ & \val{iso-srad-err-05} & \val{iso-srad-err-50} & \val{iso-srad-err-95}  & ~ & \val{iso-floor-srad-err-05}   & \val{iso-floor-srad-err-50} & \val{iso-floor-srad-err-95}  \\ 
$\sigma(\log$ age)      & \val{iso-slogage-err-05} & \val{iso-slogage-err-50} & \val{iso-slogage-err-95}  & ~ & \val{iso-floor-slogage-err-05}   & \val{iso-floor-slogage-err-50} & \val{iso-floor-slogage-err-95}  
\enddata
\tablecomments{Summary of the uncertainties associated with stellar mass, radius, and age. The ``formal'' uncertainties returned by the \isochrones framework do not incorporate model-dependent errors associated with the DSEP models. The uncertainties depend on a star's position in the HR diagram and we summarize the range of errors by quoting the 5, 50, and 95 percentiles. The ``adopted'' uncertainties incorporate an additional error terms, described in Section~\ref{ssec:uncertainties},}
\end{deluxetable*}

\subsection{Comparison with Asteroseismology}
\label{sec:as}
To verify the integrity of our derived stellar masses and radii, we performed a comparison with values computed by \cite{Huber13} (H13 hereafter) using asteroseismology for $\val{cks-huber-nstars}$ stars in common. H13 used the power in different Fourier modes in the \Kepler light curves to derive \Mstar and \Rstar with precisions of 7\% and 3\%, respectively. Aside from a weak dependence on \teff, which is determined from spectroscopy, asteroseismology relies on an independent set of observations and offers a good check on the precision and accuracy of our derived parameters. Furthermore, H13 relied on a suite of six stellar structure models%
\footnote{ASTEC \citep{ChristensenDalsgaard08}, BaSTI \citep{Pietrinferni04}, DSEP \citep{Dotter08}, Padova \citep{Marigo08}, Yonsei-Yale \citep{Demarque04}, and YREC \citep{Demarque08}.}
that reduce the risk of systematic offsets in \Mstar or \Rstar common to both H13 and CKS. 

\begin{figure*}
\gridline{
\fig{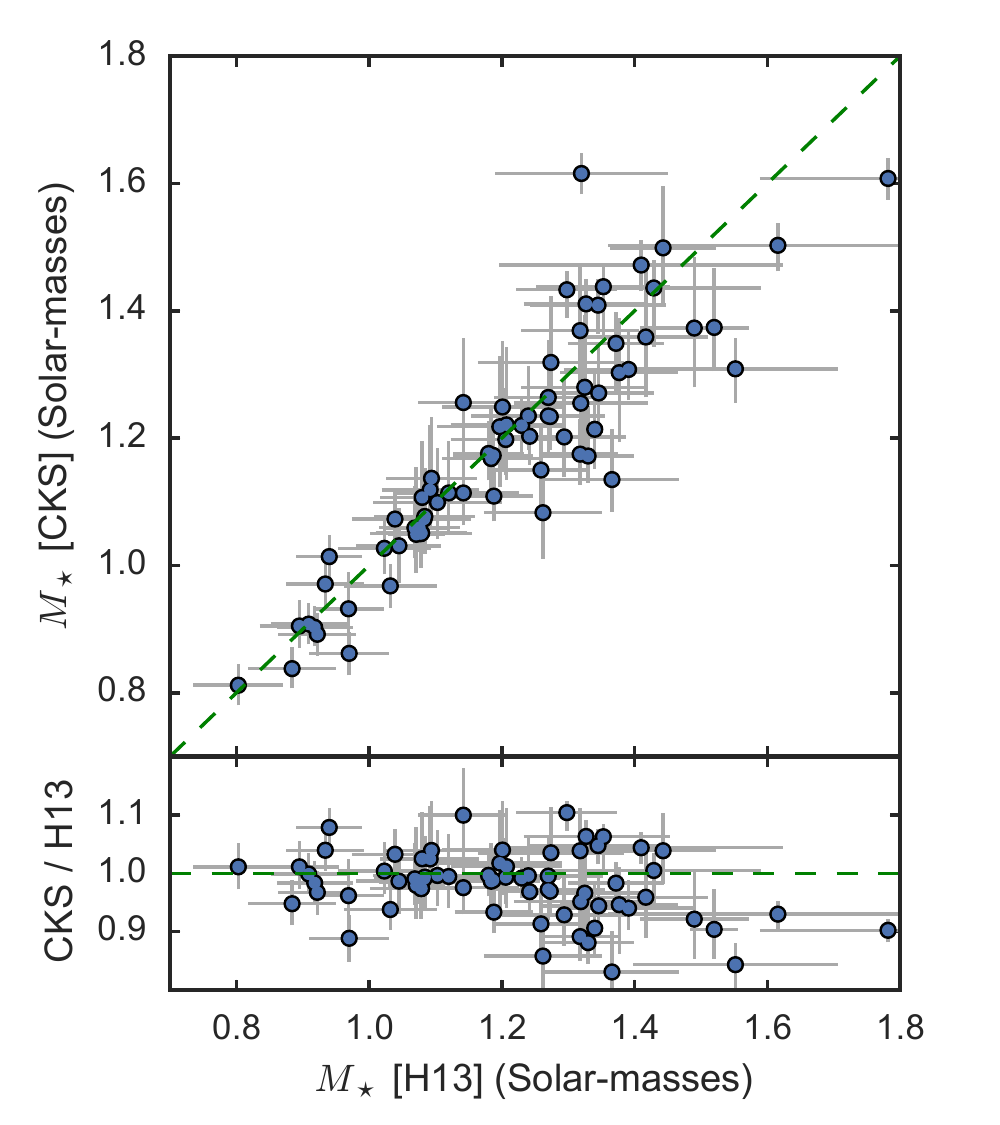}{0.5\textwidth}{}
\fig{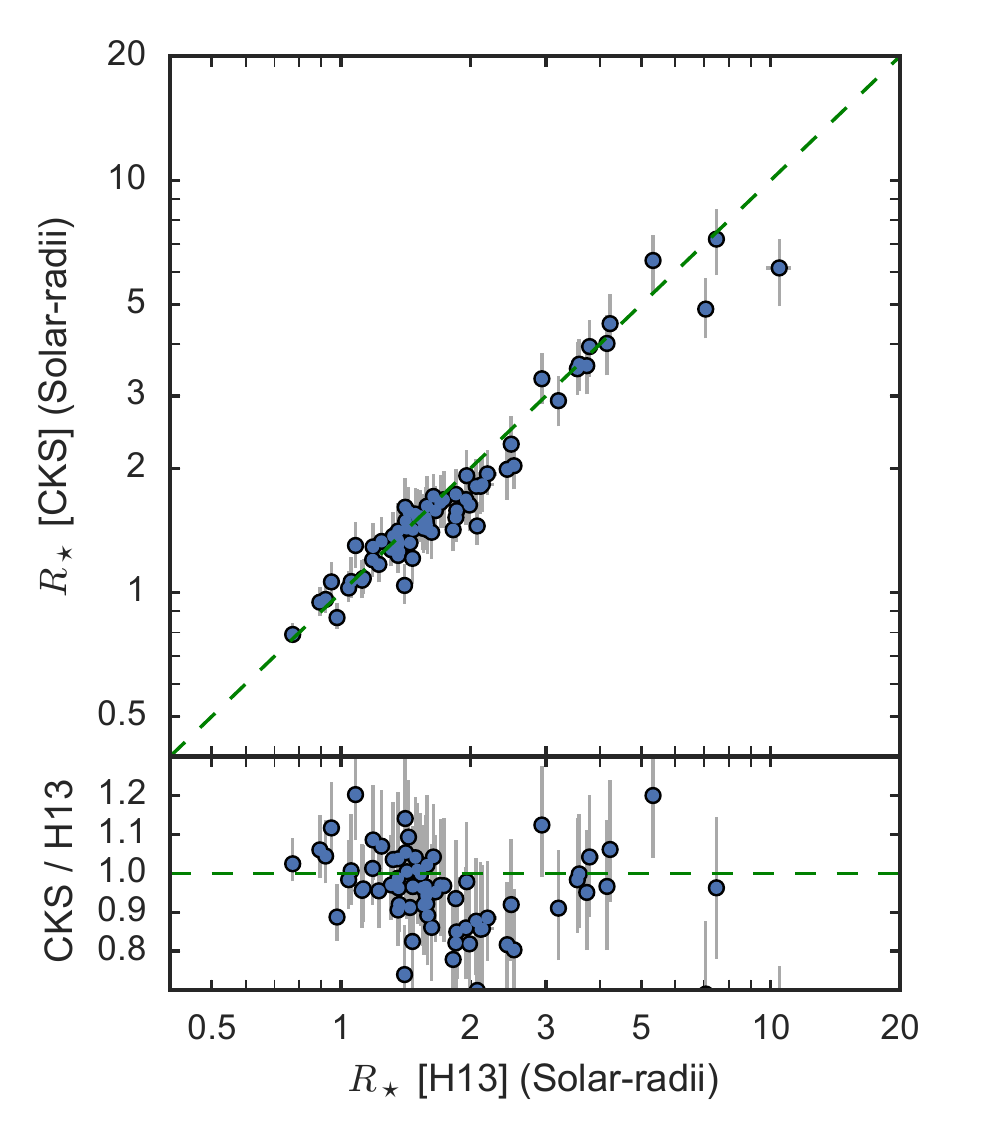}{0.5\textwidth}{}
}
\caption{Stellar masses (\Mstar) and radii (\Rstar) derived from asteroseismology (\citealt{Huber13}; H13) and spectroscopy (this work) for $\val{cks-huber-nstars}$ stars in common. {\em Left}: comparison of spectroscopic and asteroseismic \Mstar (linear scale). Equality is represented by the green line. We note that the spectroscopic \Mstar are \val{cks-huber-smass-offset} on average and that there is a $\val{cks-huber-smass-ratio-std}$ RMS dispersion in the ratios. {\em Right}: comparison of spectroscopic and asteroseismic \Rstar (log scale). For dwarf stars (94\% of the CKS sample), we find that the spectroscopic \Rstar are \val{cks-huber-dwarf-srad-offset} on average and there is a $\val{cks-huber-dwarf-srad-ratio-std}$ RMS dispersion in the ratios.\label{fig:cks-huber}}
\end{figure*}

In Figure~\ref{fig:cks-huber}, we compare \Mstar determined from spectroscopy and asteroseismology. On average, the spectroscopic \Mstar values are \val{cks-huber-smass-offset} than the asteroseismic values
%
%
with a $\val{cks-huber-smass-ratio-std}$ RMS scatter in the ratio. We assessed the degree to which the errors associated with the CKS and H13 masses can account for the observed dispersion by computing the ``reduced-chi-square'' statistic:
\[
\chi_r^2 = \frac{1}{N} \sum \frac{(M_{\star,2} - M_{\star,1})^{2}}{\sigma_1^2 + \sigma_2^2},
\]
where subscripts 1 and 2 refer to the CKS and H13 parameters, respectively. A \rchisq = 1 indicates that the quoted uncertainties can account for the observed dispersion. For the CKS-H13 \Mstar comparison, \rchisq = \val{cks-huber-smass-rchisq}, indicating reasonable errors.

Figure~\ref{fig:cks-huber} also shows the agreement between spectroscopic and asteroseismic \Rstar. The median uncertainty on the spectroscopic radii is $\val{iso-floor-srad-err-50}$, while the asteroseismic radii are measured to 3\% \citep{Huber13}. When examining both dwarf and giant stars, we observe an RMS scatter of $\val{cks-huber-srad-ratio-std}$. For dwarf stars (94\% of the CKS sample), the agreement is slightly tighter, with an RMS scatter of $\val{cks-huber-dwarf-srad-ratio-std}$. For the CKS-H13 \Rstar comparison, we find \rchisq = \val{cks-huber-srad-rchisq}, indicating that the reported uncertainties can account for the observed scatter. 


We note a systematic trend in the ratio of the CKS and H13 \Rstar in the range of 1--3~\Rsun. While one could, in principle, improve the agreement between spectroscopic and asteroseismic radii with an {\em ad hoc} correction, we elect against adding this additional complication. The observed trend may raise some concern regarding the accuracy of \Rstar for stars smaller than 1.0~\Rsun, where few asteroseismic anchor points exist. However, the spread in main sequence stellar radii rapidly shrinks toward later type stars. For the cool dwarfs in the CKS sample, the stellar radii are primarily constrained by the spectroscopic effective temperatures, which are precise to 60~K.

Spectroscopy and isochrone modeling provide some information regarding the stellar age, although this parameter is not as well-constrained as either \Mstar or \Rstar in a fractional sense. For the CKS sample, the median uncertainty is \val{iso-floor-slogage-err-50}. Here, we assess the integrity of these uncertainties with comparisons to asteroseismology. As stars evolve, nuclear fusion changes the radial distribution of stellar mass, $\rho(r)$. In some cases, the frequencies of individual oscillation modes can be measured from photometry, and asteroseismology can probe $\rho(r)$. In these cases, asteroseismology provides additional leverage on stellar age beyond \teff, \logg, and \fe. 

\begin{figure*}
\gridline{
\fig{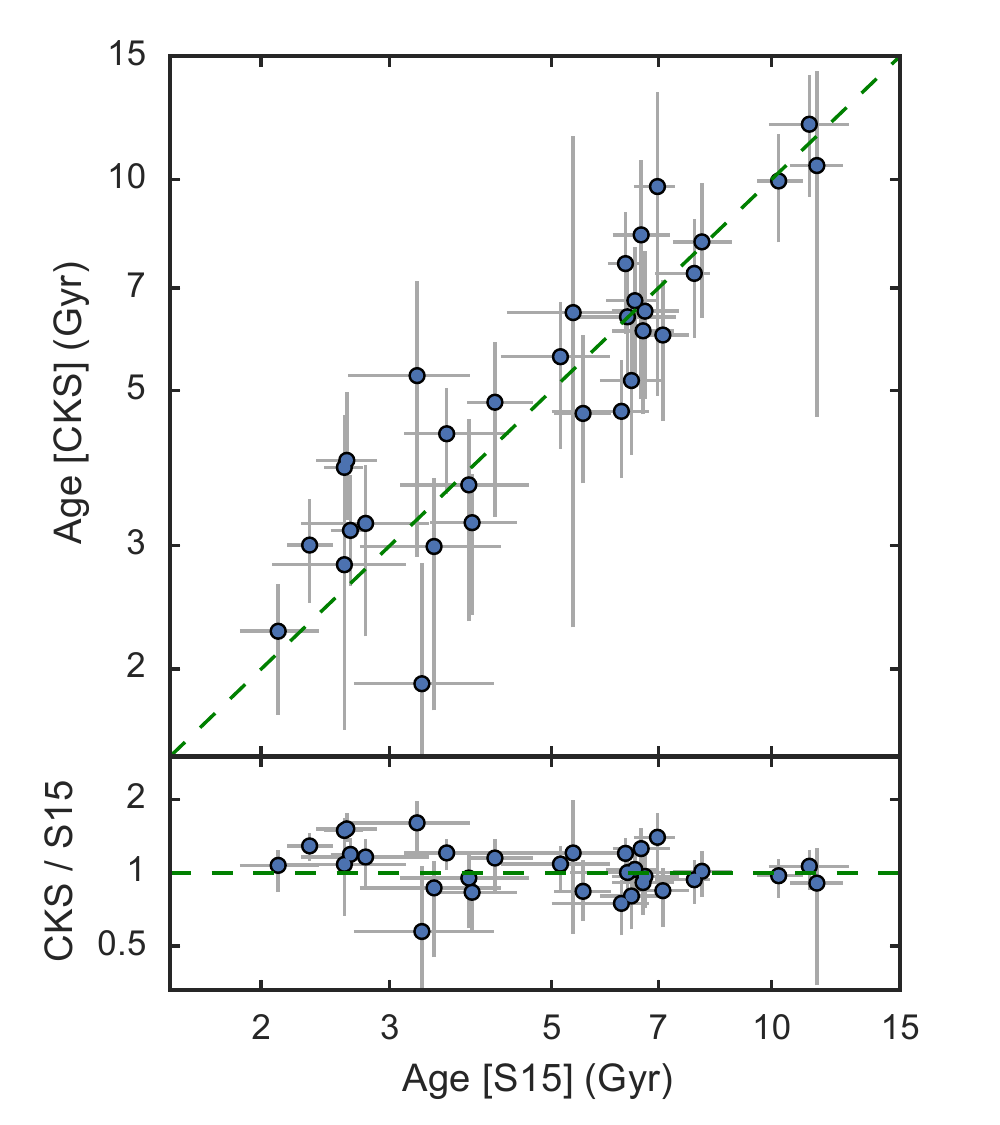}{0.5\textwidth}{}
\fig{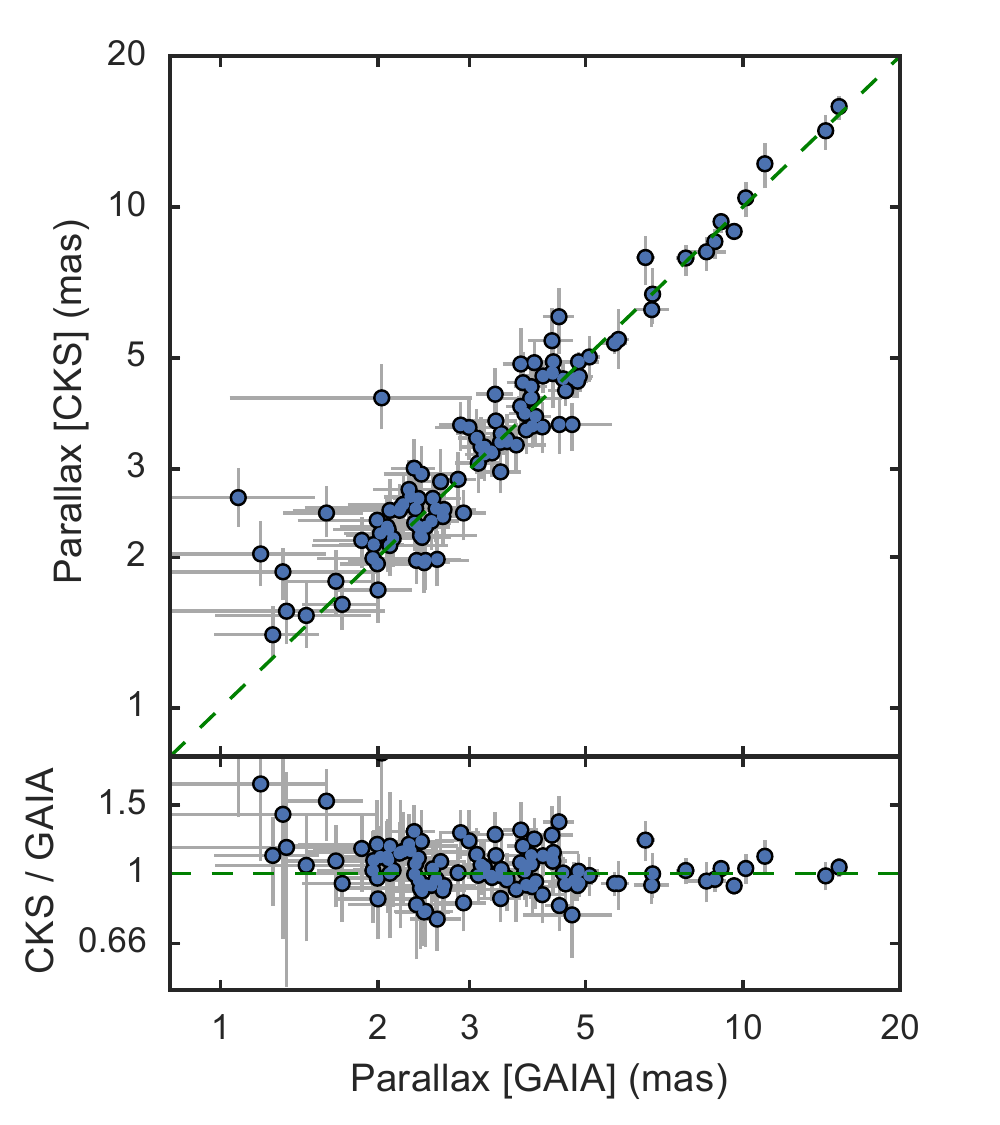}{0.5\textwidth}{}
}
\caption{{\em Left:} Comparison of stellar ages derived from asteroseismology (\citealt{Silva-Aguirre15}; S15) and spectroscopy (this work). On average, the CKS ages are \val{cks-silva-slogage-diff-offset} with a scatter of \val{cks-silva-slogage-diff-std} in the ratios. {\em Right:} Comparison stellar parallax derived from spectroscopy, isochrones, $K$-band photometry (\plxiso) and parallaxes from the Tycho-Gaia Astrometric Solution (\plxtrig). The majority of sample is consistent within errors, with a possible systematic offset for the most distant stars.\label{fig:age-parallax}}
\end{figure*}

In Figure~\ref{fig:age-parallax}, we show a comparison between CKS ages and ages from asteroseismic modeling of individual modes performed by \cite{Silva-Aguirre15} (S15 hereafter) for \val{cks-silva-nstars} stars in common. S15 report median fractional age uncertainties of 0.056~dex.  On average, the CKS ages are \val{cks-silva-slogage-diff-offset} with a scatter of \val{cks-silva-slogage-diff-std}.  For the CKS-S15 age comparison, we find \rchisq = \val{cks-silva-slogage-rchisq}, which indicates adopted errors can reasonably account for the the observed scatter. We note that the quality the CKS age constraints varies across the HR diagram. For cool dwarf stars, age is only constrained to a factor of two.

\subsection{Comparison using \Gaia Parallaxes}
\label{sec:gaia}
 While we treat asteroseismology as the preferred benchmark with which to assess our spectroscopic parameters and uncertainties, we performed an additional assessment of the quality of the derived CKS stellar radii using trigonometric parallaxes, \plxtrig, from the Tycho-Gaia Astrometric Solution (TGAS; \citealt{Gaia16a,Gaia16b,Lindegren16,Michalik2015}). As discussed in Section~\ref{ssec:isochrone}, one of the outputs of our isochrone modeling is a parallax measurement, \plxiso. Comparing the two measurements of parallax is a good check on the quality of the CKS measurements of \Rstar. For example, if the CKS stellar radii were systematically large, the inferred distance to the stars would be systematically large, resulting in measurements of \plxiso that are systematically smaller than \plxtrig.

We compare \plxiso and \plxtrig in Figure~\ref{fig:age-parallax}. On average \plxiso is \val{cks-tgas-sparallax-diff-offset} than \plxtrig. The two measurements of parallax have an RMS dispersion of \val{cks-tgas-sparallax-diff-std}~mas, consistent with the typical uncertainties. We note that for the most distant objects in this sub-sample (having $\plxtrig < 2$~mas), \plxiso is often larger than the TGAS \plxtrig. As discussed in Paper I, the CKS program achieved uniform signal-to-noise on targets brighter than $K_{p} = 14.2$, regardless of distance, i.e. all stars in common. Therefore, we consider the possibility of an onset of systematic errors in the CKS parameters at parallaxes less than 2~mas unlikely.

The possibility of systematic offsets in the TGAS distance scale has been the subject of considerable interest in the past year.  \cite{Stassun16} found that the TGAS parallaxes were on average $0.25$~mas smaller than those constrained from eclipsing binaries. Similar offsets were also reported by \cite{Jao16}, who compared TGAS parallaxes to literature values of nearby stars, and by \cite{Silva-Aguirre17}, who compared TGAS parallaxes against asteroseismic parallaxes for 66 main-sequence stars. In contrast, \cite{Huber17}, who performed a comparison of TGAS and asteroseismic parallaxes for 2200 stars, observed no offset, nor did \cite{Casertano17}, who compared TGAS parallaxes to parallaxes derived from Cepheids.

The small number of comparison stars with \plxtrig < 2~mas combined with the large fractional errors in the TGAS for such stars, prevents a detailed assessment of systematics in the TGAS. We expect that this offset will diminish in future \Gaia data releases that will rely solely on \Gaia measurements. In the near future, \Gaia will provide parallaxes for all stars in the CKS sample. Comparisons between the CKS parallaxes and \Gaia parallaxes will enable detailed assessments of systematics inherent to both CKS and \Gaia and constrain dust extinction in the direction of the \Kepler field.

\subsection{Comparison with Photometric Parameters}
We compare our new stellar parameters to those in the Q1-Q16 KOI catalogue \citep{Mullally15}, which we accessed via the NASA Exoplanet Archive \citep{Akeson13}%
\footnote{http://exoplanetarchive.ipac.caltech.edu/} 
on 2016-12-12. The Q1-Q16 KOI catalog (Q16 hereafter) contains the stellar properties of \cite{Huber14}, which were derived from various literature sources based on asteroseismology, spectroscopy, and photometry. 


The vast majority, $\val{huberphot-nstars-phot}$/$\val{nstars-cks}$ (74\%) of the stars in the \cite{Huber14} catalog that appear in the CKS sample have only photometric constraints on \logg. However, only $\val{huberphot-nstars-ast}$/$\val{nstars-cks}$ (7\%) of CKS stars had previous asteroseismic constraints, and $\val{huberphot-nstars-spec}$/$\val{nstars-cks}$ (17\%) had previous spectroscopic constraints on \logg. Our new spectroscopic constraints on \logg and stellar radius are generally more precise than the previous photometric or spectroscopic constraints, but we do not improve the stellar radius precision for stars that already had asteroseismic constraints.

Median uncertainties in the Q16 catalog are $\val{nea-smass-frac-err-median}$ and $\val{nea-srad-frac-err-median}$ for stellar mass and radius respectively, while the median uncertainties presented in this work are $\val{iso-floor-smass-err-50}$ and $\val{iso-floor-srad-err-50}$ for stellar mass and radius respectively. We computed the fractional differences in stellar radii, 
\[
\frac{\Delta \Rstar}{\Rstar} = \frac{R_{\star,\mathrm{CKS}}- R_{\star,\mathrm{Q16}}}{R_{\star,\mathrm{CKS}}},
\]
to assess the offset and scatter between the two samples. When considering all CKS stars, we found a modest offset between the CKS and Q16 radii, mean($\Delta\Rstar/\Rstar$)~=~\val{cks-huberphot-srad-ratio-mean} and a scatter of RMS($\Delta \Rstar / \Rstar$)~=~\val{cks-huberphot-srad-ratio-std} after removing $\val{cks-huberphot-srad-noutliers}$ outliers with radii differing by more than a factor of two. We computed the fractional differences in stellar masses,
\[
\frac{\Delta \Mstar}{\Mstar} = \frac{M_{\star,\mathrm{CKS}}- M_{\star,\mathrm{Q16}}}{M_{\star,\mathrm{CKS}}}.
\]
On average, the CKS masses had a small offset with respect to the Q16 masses, mean($\Delta \Mstar / \Mstar$) = $\val{cks-huberphot-smass-ratio-mean}$ with a scatter RMS($\Delta \Mstar / \Mstar$) = $\val{cks-huberphot-smass-ratio-std}$ after removing $\val{cks-huberphot-smass-noutliers}$ outliers with masses differing by more than a factor of two. 

\begin{figure*}
\gridline{\fig{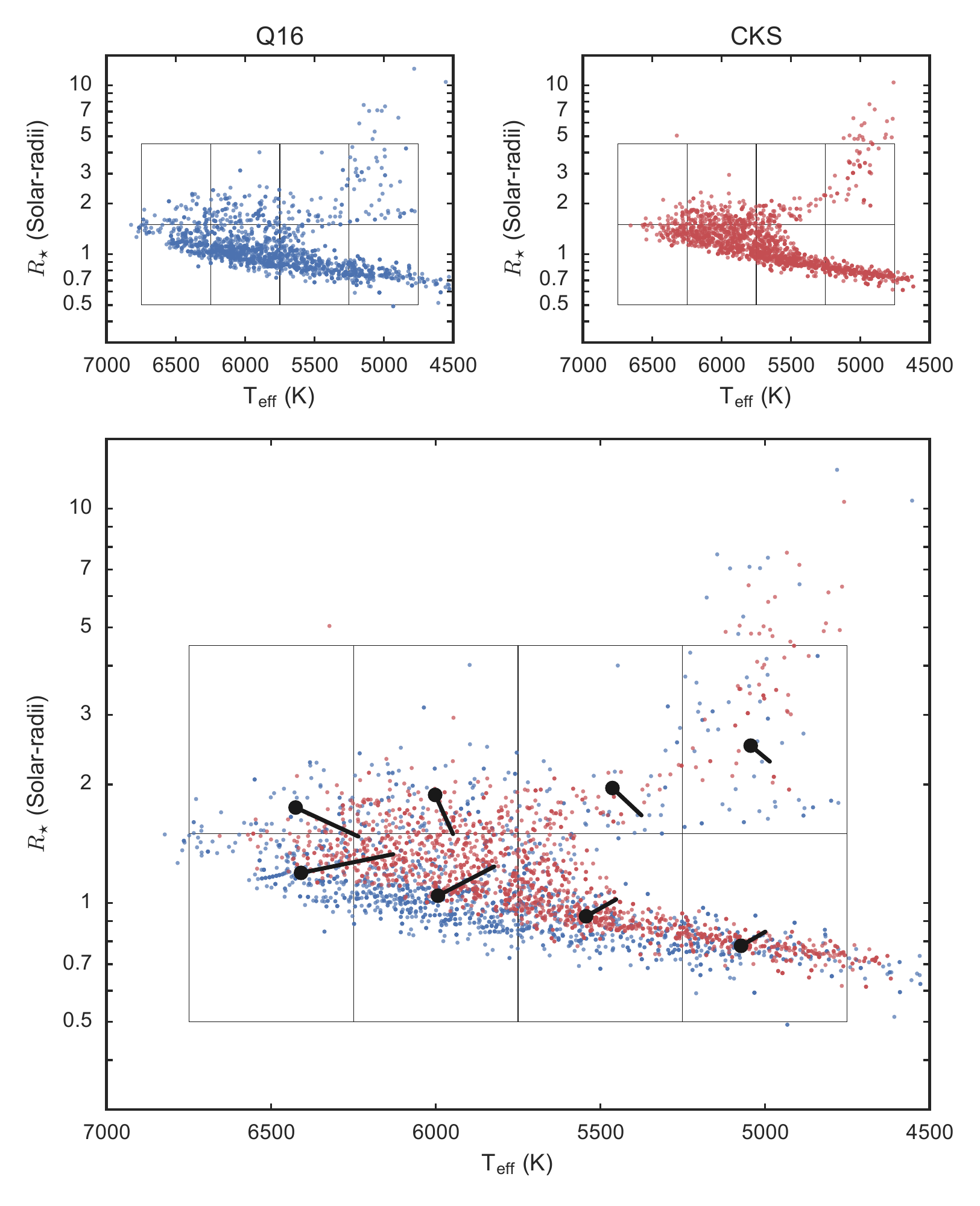}{0.85\textwidth}{}}
\caption{{\em Top left}: \Rstar and \teff from the Q1-Q16 KOI Catalog (Q16; \citealt{Mullally15}) for the stars in the CKS sample. The parameters are primarily based on broadband photometry, with a small number from astroseismology and previous spectroscopic studies. The green bar reflects the median uncertainties. {\em Top right}: Same but showing spectroscopic parameters from this work. {\em Bottom}: Enlarged representation of the CKS and Q16 parameters to highlight differences between the two samples. We identify stars having Q16 properties that fall within each of the black boxes and the circles represent the mean Q16 (\teff, \Rstar). The lines point to the mean CKS (\teff, \Rstar) for these same stars to highlight the systematic offsets in between two catalogs as a function of \teff and \Rstar. The largest difference is for the hottest stars which have systematically lower spectroscopic temperatures. A number of stars that the Q16 catalog designates as subgiants are reclassified as dwarfs which account for the downward shift in the upper right grid cell.}
\label{fig:teff-radius}
\end{figure*}

We compare the Q16 and CKS radii as a function of effective temperature in Figure~\ref{fig:teff-radius}. Although the average CKS and Q16 radii agree at the few percent level, we note significant temperature-dependent systematics for stars having $\teff \gtrsim 6000$~K. For dwarf stars ($\Rp < 1.5~\Rsun$), the CKS parameters prefer cooler and slightly larger stars. For slightly-evolved stars ($\Rp > 1.5~\Rsun$) the CKS stellar properties favor cooler and smaller stars. For the hottest stars the typical offset between the spectroscopic and photometric \teff reaches 200~K.

Measuring \teff and \logg from photometry introduces systematics, which are discussed in previous stellar classification papers (e.g. \citealt{Pinsonneault12,Huber14}). These systematics are due to the fact that photometry provides little independent leverage on \teff, \logg, and reddening. Both \cite{Pinsonneault12} and \cite{Huber14} apply {\em ad hoc} corrections to the photometric \teff which grow to 400~K at 6500~K. Given that we observe offsets of 200~K, we conclude that these {\em ad hoc} corrections did not completely remove the systematic errors associated with photometric \teff.

\clearpage

\section{Planet Properties}
\label{sec:planet}
We used our newly-measured stellar parameters to re-calculate several important planetary parameters. We began with the transit fit parameters from the Q16 KOI catalog \citep{Mullally15}. We re-computed planet radii (\Rp) using the published transit depths  and the CKS \Rstar. Given that the planet radii are limited by uncertainties in the stellar radii, the CKS stellar-radii enable an improvement of planet radii \Rp from $\sigma(\Rp) /\Rp \approx \val{nea-rp-frac-err-median}$ to $\val{cks-rp-frac-err-median}$.

\begin{figure*}
\label{fig:radius-dist}
\gridline{
\fig{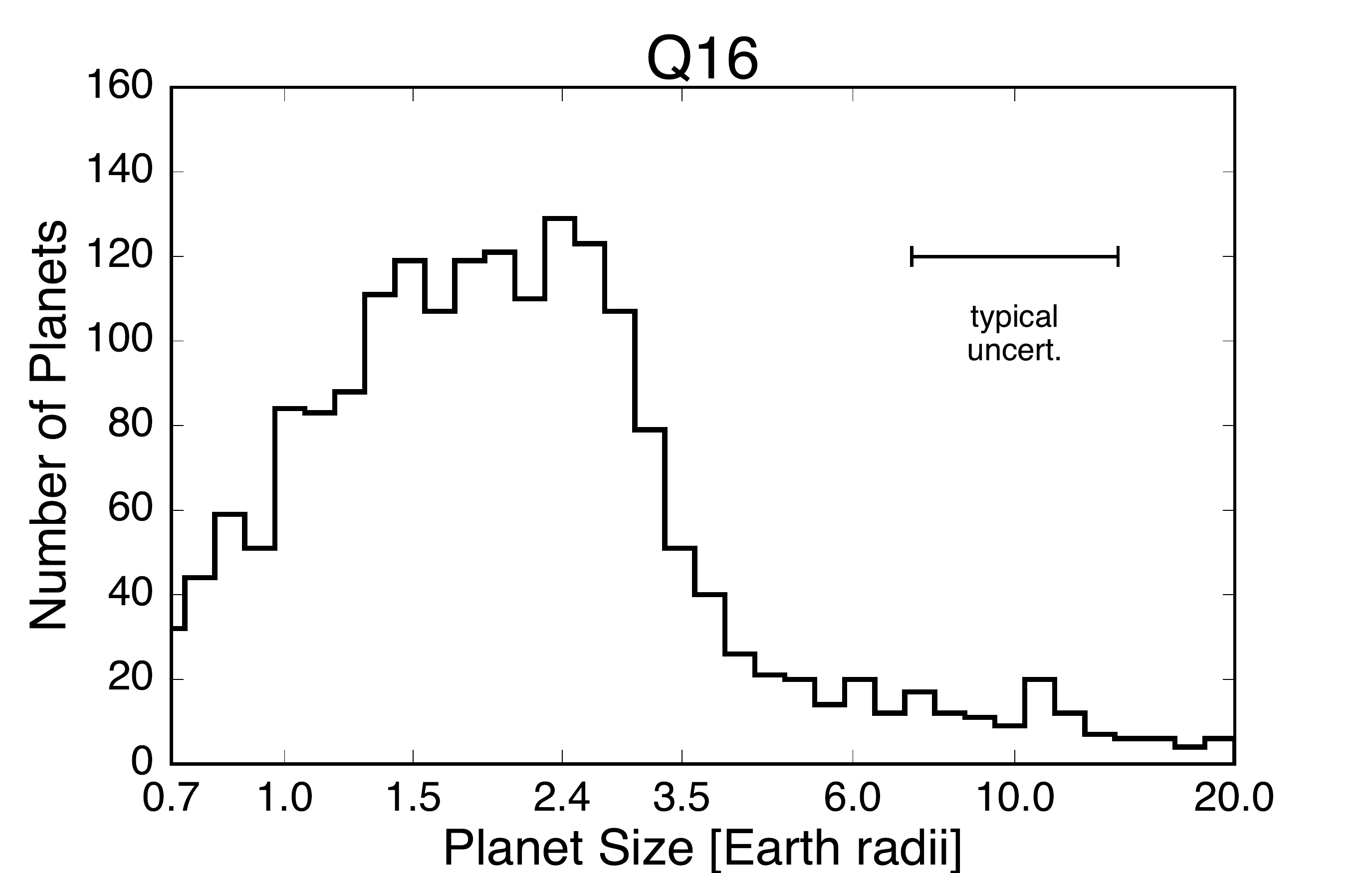}{0.5\textwidth}{}
\fig{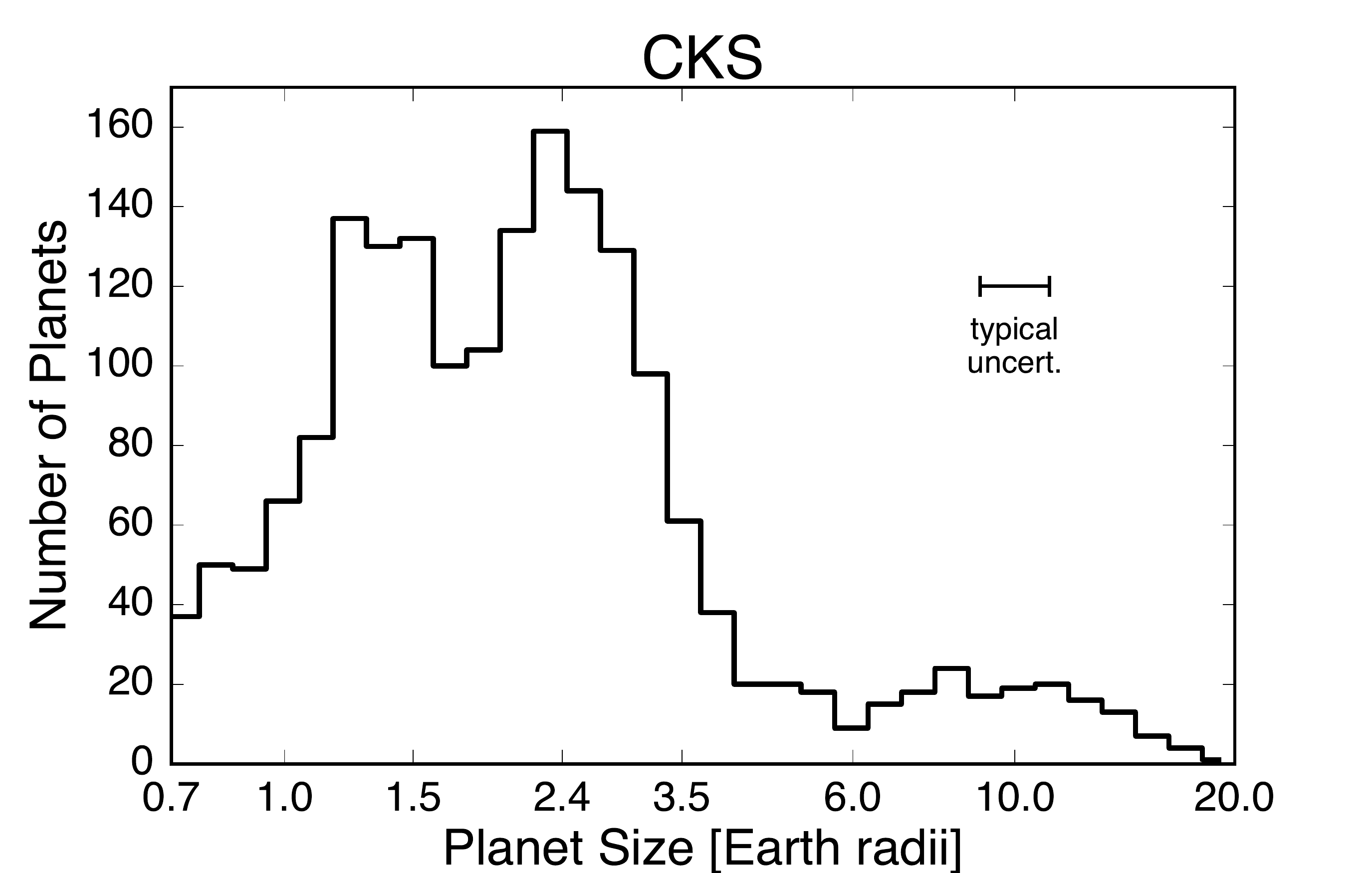}{0.5\textwidth}{}
}
\caption{\emph{Left:} Number of CKS planet candidates having different sizes. Here, the planet radii are taken from the Q1-Q16 KOI catalog (Q16; \citealt{Mullally15}). The error bar shows the median uncertainty in planet radius. \emph{Right:} Same but showing planet radii computed using the CKS spectroscopic parameters. We note the emergence of structure in the CKS histogram of radii, the statistical significance of which requires further work, presented in Fulton et al. (2017, submitted).}
\end{figure*}

Figure~\ref{fig:radius-dist} shows the distribution of planet radii from the Q16 catalog and from this paper. The general features of the two histograms are similar. We note apparent structure in the histogram of the CKS radii that is not apparent in the Q16 histogram.  The statistical significance of this structure in the planet radius distribution will be explored in detail in Paper III of this series (Fulton et al. 2017, submitted).

Using our updated stellar properties, we recomputed planet semi-major axes $a$, incident stellar flux \Sinc, and equilibrium temperature \teq, assuming circular orbits.  Semi-major axes are computed using Kepler's Third Law. Because orbital periods are measured very precisely from \Kepler photometry, uncertainty in $a$ is set by the uncertainty in \Mstar according to
%
%
\begin{eqnarray}
\frac{\sigma(a)}{a} =  \frac{1}{3} \frac{\sigma(\Mstar)}{\Mstar} \approx  1.7\%.
\end{eqnarray}
One could, in principle, compute $a$ from $\Rstar/a$, measured from the transit profile, and $\Rstar$. However, $\Rstar/a$ has large uncertainties due to the assumption of a circular orbits and degeneracies with impact parameter. At a minimum, the uncertainty in semi-major axis computed according to this second method is 
\begin{equation}
\frac{\sigma(a)}{a} \gtrsim \frac{\sigma(\Rstar)}{\Rstar} \approx 10\%,
\end{equation}
which is far less precise than the calculation using Kepler's Third Law.
 
We compute the incident flux as
\begin{equation}
\frac{\Sinc}{S_{\oplus}} = \left(\frac{\teff}{5778\, \mathrm{K}}\right)^4 
\left(\frac{\Rstar}{\Rsun}\right)^2
\left(\frac{a}{\, \mathrm{AU}}\right)^{-2}.
\end{equation}
For convenience, we also provide planetary equilibrium temperature, \teq, defined according to
\begin{equation}
\left(\frac{\teq}{280~\mathrm{K}}\right) = \left(\frac{\Sinc}{S_{\oplus}}\right)^{1/4}
\left(\frac{1-\alpha}{4}\right)^{1/4},
\end{equation}
assuming a Bond albedo $\alpha$ of 0.3, typical for super-Earth-size planets \citep{Demory14}. Because \Sinc depends on $\teff^4$ and $\Rstar^2$ our spectroscopic improvements \teff and \Rstar result in a substantial improvement in \Sinc from $\sigma(\Sinc)/\Sinc$ of \val{nea-sinc-frac-err-median} to \val{cks-sinc-frac-err-median}. The updated planetary parameters for the $\val{ncand-cks}$ planet candidates in the CKS sample are listed in Table~\ref{tab:planet-parameters}.

\begin{deluxetable*}{lrrrrrr}
\tablecaption{Summary of Typical Parameter Uncertainties\label{tab:summary}}
\tablecolumns{7}
\tablewidth{0pt}
\tablehead{
\colhead{Source} &	
\multicolumn{4}{c}{Q16}  &
\colhead{} &
\colhead{CKS} \\
\cline{2-5} 
\cline{7-7} \\[-3.5ex]
\colhead{Method} & 
\colhead{All} &
\colhead{AS} & 
\colhead{Spec.} &
\colhead{Phot.} &
\colhead{} &
\colhead{Spec.} \\
\colhead{$N_\star$} & 
\colhead{\val{huberphot-nstars-tot}} &
\colhead{\val{huberphot-nstars-ast}} & 
\colhead{\val{huberphot-nstars-spec}} & 
\colhead{\val{huberphot-nstars-phot}} &
\colhead{} &
\colhead{{\bf \val{nstars-cks}}}
}
\startdata
$\sigma(\Mstar)/\Mstar$  & \val{huberphot-smass-frac-err-median}  & \val{huberphot-smass-ast-frac-err-median} & \val{huberphot-smass-spec-frac-err-median} & \val{huberphot-smass-phot-frac-err-median} & & {\bf \val{iso-floor-smass-err-50}}  \\ 
$\sigma(\Rstar)/\Rstar$   & \val{huberphot-srad-frac-err-median}  & \val{huberphot-srad-ast-frac-err-median} & \val{huberphot-srad-spec-frac-err-median} & \val{huberphot-srad-phot-frac-err-median} & & {\bf \val{iso-floor-srad-err-50}}   \\
$\sigma(\log$ age)      & \nodata   &  \nodata  & \nodata   &  \nodata  &  &{\bf \val{iso-floor-slogage-err-50}} \\ 
$\sigma(\Rp)/\Rp$         & \val{nea-rp-frac-err-median}  &  \val{nea-rp-frac-err-median-ast}  &  \val{nea-rp-frac-err-median-spec}  & \val{nea-rp-frac-err-median-phot}   & & {\bf \val{cks-rp-frac-err-median}}  \\
$\sigma(\Sinc)/\Sinc$     & \val{nea-sinc-frac-err-median}   &  \val{nea-sinc-frac-err-median-ast}  & \val{nea-sinc-frac-err-median-spec}   &  \val{nea-sinc-frac-err-median-phot}   &&  {\bf \val{cks-sinc-frac-err-median}}  \\
$\sigma(a)/a$             &  \nodata                          &  \nodata                           &  \nodata                               &  \nodata                                && {\bf \val{cks-sma-frac-err-median}}  
\enddata
\tablecomments{Summary of median quoted uncertainties for Q1-Q16 KOI catalog (Q16; \citealt{Mullally15}) and the CKS sample. We also list uncertainties for the sub-samples of the Q16 parameters based on asteroseismology, spectroscopy, or photometry. The CKS survey contains a few dozen stars not included in the Q16 catalog.}
\end{deluxetable*}

\section{Conclusion}
\label{sec:conclusion}
In this work, we converted the measured spectroscopic stellar parameters presented in Paper~I to the physical stellar masses, radii, and ages for $\val{nstars-cks}$ stars in the CKS sample. We used these properties to improve knowledge of the physical properties of $\val{ncand-cks}$ planet candidates including \Rp and \Sinc. 

These improved stellar and planet properties will yield new insights into the \Kepler sample of planets some of which will be explored in subsequent papers in this series. Paper~III (Fulton et al. 2017) examines the planet radius distribution, brought into sharper focus by the improved uncertainties in planet size. In Paper~IV (Petigura et al. 2017) we explore the extent to which host star metallicity is connected to other planet properties. Paper~V (Weiss et al. 2017), explores the connection between stellar and planet properties in the context of planetary multiplicity and system architectures. 

Finally, we encourage the community to use the CKS dataset. All stellar spectra analyzed here are available to the public via the Keck Observatory Archive,%
\footnote{\url{http://www2.keck.hawaii.edu/koa/public/koa.php}}
the Community Follow-up Program (CFOP) website,%
\footnote{\url{http://cfop.ipac.caltech.edu}}
and the CKS project website.%
\footnote{\url{http://astro.caltech.edu/~howard/cks/}}
The CFOP website also contains additional information about each KOI and a discussion of the available follow-up observations. The spectroscopic and derived stellar parameters are available from the CFOP and the CKS project website. The code used to produce the derived parameters is available on GitHub.%
\footnote{\url{https://github.com/California-Planet-Search/cksphys/} (v3.0)}
We expect and anticipate that these data will prove useful for many additional projects.

\facilities{Keck:I (HIRES),  Kepler}

\acknowledgments{
The CKS project was conceived, planned, and initiated by AWH, GWM, JAJ, HTI, and TDM.  
AWH, GWM, JAJ acquired Keck telescope time to conduct the magnitude-limited survey.  
Keck time for the other stellar samples was acquired by JNW, LAR, and GWM.
The observations were coordinated by HTI and AWH and carried out by AWH, HTI, GWM, JAJ, TDM, BJF, LMW, EAP, ES, and LAH. 
AWH secured CKS project funding. 
SpecMatch was developed and run by EAP and SME@XSEDE was developed and run by LH and PAC. 
EAP computed derived planetary and stellar properties with assistance from BJF.
This manuscript was largely written by EAP with significant assistance from AWH, GWM, and BJF.

We thank Jason Rowe, Dan Huber, and Jeff Valenti for helpful conversations and Roberto Sanchis-Ojeda for his work on the Ultra-Short Period planet sample.
We thank the many observers who contributed to the measurements reported here. 
We gratefully acknowledge the efforts and dedication of the Keck Observatory staff, especially Randy Campbell, Scott Dahm, Greg Doppmann, Marc Kassis, Jim Lyke, Hien Tran, Josh Walawender, Greg Wirth for support of HIRES and of remote observing. Most of the data presented here are based on spectra obtained at the W.\ M.\ Keck Observatory, which is operated as a scientific partnership among the California Institute of Technology, the University of California, and NASA. We are grateful to the time assignment committees of the University of Hawaii, the University of California, the California Institute of Technology, and NASA for their generous allocations of observing time that enabled this large project.
Kepler was competitively selected as the tenth NASA Discovery mission. Funding for this mission is provided by the NASA Science Mission Directorate.  
We thank the , the Kepler Science Office, the Science Operations Center, Threshold Crossing Event Review Team (TCERT), and the Followup Observations Program (FOP) Working Group for their work on all steps in the planet discovery process ranging from selecting target stars and pointing the \Kepler telescope to developing and running the photometric pipeline to curating and refining the catalogs of \Kepler planets.  
EAP acknowledges support from Hubble Fellowship grant HST-HF2-51365.001-A awarded by the Space Telescope Science Institute, which is operated by the Association of Universities for Research in Astronomy, Inc. for NASA under contract NAS 5-26555. 
AWH acknowledges NASA grant NNX12AJ23G. 
TDM acknowledges NASA grant NNX14AE11G.
PAC acknowledges NSF grant AST-1109612.
LH acknowledges NSF grant AST-1009810.
LMW acknowledges support from Gloria and Ken Levy and from the the Trottier Family.
ES is supported by a post-graduate scholarship from the Natural Sciences and Engineering Research Council of Canada.
This work made use of the SIMBAD database (operated at CDS, Strasbourg, France), NASA's Astrophysics Data System Bibliographic Services, and the NASA Exoplanet Archive, which is operated by the California Institute of Technology, under contract with the National Aeronautics and Space Administration under the Exoplanet Exploration Program.
This work has made use of data from the European Space Agency (ESA) mission Gaia, processed by the Gaia Data Processing and Analysis Consortium. Funding for the DPAC has been provided by national institutions, in particular the institutions participating in the Gaia Multilateral Agreement.

Finally, the authors wish to recognize and acknowledge the very significant cultural role and reverence that the summit of Maunakea has always had within the indigenous Hawaiian community.  We are most fortunate to have the opportunity to conduct observations from this mountain.}

\onecolumngrid
\begin{deluxetable*}{lRRRRRr}
\tablecaption{CKS Planet Parameters\label{tab:planet-parameters}}
\tabletypesize{\scriptsize}
\tablecolumns{6}
\tablewidth{0pt}
\tablehead{
	\colhead{Planet} & 
	\colhead{$P$\tablenotemark{1}} &
	\colhead{\Rp/\Rstar\tablenotemark{1}} & 
	\colhead{\Rp} & 
	\colhead{\Sinc\tablenotemark{2}} &
	\colhead{\teq \tablenotemark{3}} & 
    \\
    \colhead{candidate} & 
	\colhead{d} &
	\colhead{} &
	\colhead{\Re} & 
	\colhead{F$_{\oplus}$} &
	\colhead{K} & 
}
\startdata
K00001.01  &  2.47  &  0.123851^{+0.000025}_{-0.000076}  &  14.32^{+1.42}_{-1.42}   &  891^{+185}_{-185}  &   1392^{+72}_{-72} \\
K00002.01  &  2.20  &  0.075408^{+0.000008}_{-0.000007}  &  13.41^{+2.02}_{-2.02}   &  3030^{+931}_{-931}  &   1891^{+146}_{-146} \\
K00003.01  &  4.89  &  0.057989^{+0.000049}_{-0.000033}  &  5.11^{+0.41}_{-0.41}   &  117^{+20}_{-20}  &   838^{+36}_{-36} \\
K00006.01  &  1.33  &  0.294016^{+0.103683}_{-0.209459}  &  39.73^{+21.97}_{-21.97}   &  3595^{+694}_{-694}  &   1973^{+95}_{-95} \\
K00007.01  &  3.21  &  0.024735^{+0.000141}_{-0.000076}  &  4.13^{+0.60}_{-0.60}   &  1234^{+367}_{-367}  &   1510^{+113}_{-113} \\
K00008.01  &  1.16  &  0.018559^{+0.000246}_{-0.001678}  &  1.99^{+0.19}_{-0.19}   &  2211^{+378}_{-378}  &   1748^{+74}_{-74} \\
K00010.01  &  3.52  &  0.093582^{+0.000117}_{-0.000198}  &  13.39^{+1.85}_{-1.85}   &  1009^{+286}_{-286}  &   1436^{+102}_{-102} \\
K00017.01  &  3.23  &  0.095137^{+0.000020}_{-0.000018}  &  15.04^{+2.10}_{-2.10}   &  979^{+284}_{-284}  &   1425^{+103}_{-103} \\
K00018.01  &  3.55  &  0.080126^{+0.000022}_{-0.000020}  &  13.94^{+2.19}_{-2.19}   &  1483^{+478}_{-478}  &   1582^{+128}_{-128} \\
K00020.01  &  4.44  &  0.117936^{+0.000016}_{-0.000023}  &  21.41^{+3.13}_{-3.13}   &  1004^{+302}_{-302}  &   1435^{+107}_{-107} \\

\enddata
\tablecomments{Table \ref{tab:planet-parameters} is available in its entirety in machine-readable format. A portion is shown here for guidance regarding its form and content.}
\tablenotetext{1}{Value from the NASA's Exoplanet Archive Q1-Q16 KOI catalogue \citep{Mullally15}.}
\tablenotetext{2}{Stellar irradiance received at the planet relative to the Earth.}
\tablenotetext{3}{Equilibrium temperature assuming a Bond albedo of 0.3 \citep{Demory14}}
\end{deluxetable*}

\bibliography{manuscript.bib}
\end{document}